\documentclass[fleqn,twoside]{article}%
\usepackage{amsmath}
\usepackage{amsfonts}
\usepackage{amssymb}
\usepackage{graphicx}%
\def\be{\begin{equation}}

\def\ee{\end{equation}}

\def\bi{\bibitem}

\begin{document}

\title{Field independent cosmic evolution}

\author{Nayem Sk.$^\dag$ and Abhik Kumar Sanyal$^{\S}$}
\maketitle

\noindent

\begin{center}
$\dag$Dept. of Physics, University of Kalyani, Nadia, India - 741235\\
$\S$Dept. of Physics, Jangipur College, Murshidabad, India - 742213\\

\end{center}

\footnotetext[1]{
Electronic address:\\

\noindent $\dag$nayemsk1981@gmail.com\\
\noindent $^{\S}$sanyal\_ ak@yahoo.com\\}

\begin{abstract}

\noindent It has been shown earlier that Noether symmetry does not admit a form of F(R) corresponding to an action in which F(R) is coupled to scalar-tensor action for gravity or even for pure F(R) gravity taking anisotropic model into account. Here, we prove that F(R) theory of gravity does not admit Noether symmetry even if it is coupled to Tachyonic field and considering a gauge in addition. To handle such a theory, a general conserved current has been constructed under a condition which decouples higher order curvature part from the field part. This condition in principle, solves for the scale factor, independently. Thus cosmological evolution remains independent of the form of the chosen field, whether it be a scalar or a tachyon.
\end{abstract}

\section{Introduction}

Interest in $F(R)$ theory of gravity has increased predominantly in recent years since it appears to explain most of the presently available cosmological data unifying early inflation with late time cosmic acceleration (see \cite{a1} and \cite{a2} for recent reviews and also references therein). However, most of these interesting results are the outcome of scalar tensor equivalence under some arbitrary choice of the form of $F(R)$. It is therefore important to test if the same results are obtainable from $F(R)$ theory of gravity without invoking scalar tensor equivalence. But then, how to choose a form of $F(R)$ out of indefinitely large number of curvature invariant terms and to find exact solutions are big questions. From physical ground viz., to obtain a renormalizable theory of gravity, a form of $F(R) = \alpha R + \beta R^2 + \gamma R_{\mu\nu}R^{\mu\nu}$ had been found in the context of early universe, which contains ghosts \cite{re}. A ghost free action has also been presented in recent years \cite{gf1, gf2}. Likewise, the only physically meaningful technique to obtain a form of $F(R)$ that migght explain late time cosmological evolution is to invoke Noether symmetry as a selection rule. This requires canonical formulation and for a general $F(R)$ theory of gravity it is only possible treating $R$ as an auxiliary variable, provided $F''(R) \ne 0$ (here, dash represents derivative with respect to $R$). In the process, it is possible to construct a point Lagrangian, and one can demand Noether symmetry to find a suitable form of $F(R)$. Following this technique, several authors \cite{b1, b2, b3, b4, b5} have found $F(R) \propto R^{\frac{3}{2}}$ in the Robertson-Walker metric both in vacuum $(\rho = p = 0)$ and pressureless dust $(p = 0)$. Although such a form of $F(R)$ shows accelerating expansion in the matter dominated era ($p = 0$), nevertheless, early decelerating phase tracks as $a \propto t^{\frac{1}{2}}$ in the matter dominated era instead of usual $t^{\frac{2}{3}}$ and $a \propto t^{\frac{4}{3}}$ in the radiation dominated era ($p = \frac{\rho}{3}$) instead of usual $t^{\frac{1}{2}}$, creating problem in explaining structure formation \cite{c}. Thus $R^{\frac{3}{2}}$ alone, in the absence of a linear term in the action, does not worth explaining presently available cosmological data \cite{c}. It has been also noticed \cite{c} that instead of the scale factor $a$, if one would have started with the basic variable $h_{ij} = z = a^2$, then $z$ becomes cyclic both in vacuum and matter dominated era and therefore such symmetry is independent of the choice of configuration space variables. Thus Noether symmetry obtained in the process is in-built and trivial. Situation could have been improved if Noether symmetry would allow linear term $R$ in the action, but it does not. To obtain a better form of $F(R)$, scalar field has been incorporated both minimally and non-minimally, but even then Noether symmetry remains absent \cite{d}.\\

Despite the fact that Noether symmetry yields nothing other than $F(R) \propto R^{\frac{3}{2}}$ in vacuum or in matter dominated era and that too only in isotropic space-time, some authors have recently claimed \cite{h} to have obtained arbitrary form of $F(R)$ taking into account a gauge term in the Noether theory. Note that the theory of gravity is very special in the context that it admits reparametrization invariance leading to the $(^0_0)$ equation of Einstein or the so called Hamiltonian constraint equation $H = 0$. This reparametrization invariance is not reflected in Noether equations. Thus, solution obtained from Noether equations, viz., the conserved current must satisfy the Hamilton constraint equation. It has been shown that the result \cite{h} is not true, since neither all the Noether equations nor the Hamilton constraint equation is satisfied for such symmetry \cite{i}. More recently, there is yet another claim \cite{j} that gauge Noether symmetry yields $F(R) \propto R^2$ taking tachyon field into account. It is important to review this fact since in our earlier works \cite{c, d} tachyon field had not been accounted for. \\

Rolling tachyon condensates originated from string theories and have interesting cosmological consequences \cite{t1} particularly because its equation of state parameter ($w=  \frac{p_T}{\rho_T} = -[1-\dot\phi^2], \rho_T = \frac {V(\phi)}{\sqrt{1-\dot\phi^2}}, p_T= -V(\phi){\sqrt{1-\dot\phi^2}}$) smoothly interpolates between $0$ and $-1$ \cite{t2}. As a result, it behaves like pressureless dust and cosmological constant in the limits. Such a nice feature initiated to construct viable cosmological models treating tachyon as the inflaton field by coupling it minimally to the gravitational field taking different self-interacting potential densities, in the form of power law, exponential and hyperbolic functions of the tachyon field \cite{t3}. Nevertheless, such models have been found to suffer from serious disease associated with density perturbations and reheating \cite{t4}. Tachyon fields with such kinds of potential densities were also used in order to describe the present accelerating period of the universe, where it behaves as dark energy \cite{t5}. Further, it has been observed that a tachyon field with an exponential potential plays the role of inflaton in the early universe and dark energy at the late \cite{t6}. Thus it appears to be a good candidate to explain late time cosmological evolution. Nonetheless, present observations suggest that the state parameter might even cross the phantom divide line $w < -1$, which is not realizable under minimal coupling. Such crossing may be possible if the tachyon field is non-minimally coupled to the gravitational field, which already exists in the literature \cite{t7, t8}. Another possibility appears if the tachyon field is coupled to $F(R)$ theory of gravity, which is our present concern. We therefore, in following sections, proceed to check if $F(R)$ theory of gravity being coupled to tachyon field admits Noether symmetry. The result we find is null, since the conserved current thus obtained does not satisfy the Hamilton constraint equation. This completes our `Tour de Noether symmetry' of $F(R)$ theory of gravity, which now states that $F(R)$ theory of gravity only admits the trivial symmetry $F(R) \propto R^{\frac{3}{2}}$ in vacuum or matter dominated era and that too in isotropic space-time only.\\

In view of the above discussion, we organize the present work in the following manner. In section 2, we review gauge Noether symmetry of $F(R)$ theory of gravity being coupled to tachyonic field to show that indeed such symmetry is not admissible. In section 3, we take up a more general action and attempt Noether symmetry once again but in vain. At this end we would like to mention that on the contrary, starting from $F(R) = R^2$, a-priori, Noether symmetry is obtainable for a scalar-tensor theory of gravity \cite{e} which we have not been able to recover. This is important, since $R^2$ is special in the sense that the corresponding action is scale invariance which is also somewhat related to Noether symmetry \cite{sd}. In section 4, we briefly enunciate our view in this regard. Section 5 is devoted to construct a general conserved current and to explain its utility in extracting solutions choosing arbitrary form of $F(R)$. In view of such conserved current, we show that the form of the field in no way dictates the cosmological evolution, rather it depends only on the form of $F(R)$ chosen. Finally, in section 6, we find the constraint to recover Newtonian limit for such an action under weak field approximation. Section 7 concludes the present work.\\

\section{Action, field equations and Noether equation}

Among all the dynamical symmetries, transformations that map solutions of the equations of motion into solutions, one can single out Noether symmetries as the continuous transformations that leave the action invariant - except for boundary terms. In formal language, Noether symmetry states that, for any regular system if there exists a vector field $X^{(1)}$, such that,
\be \left(\pounds _{X^{(1)}} + \frac{d\eta}{dt}\right)L = \left(X^{(1)} + \frac{d\eta}{dt}\right) L = \frac{d B}{d t},\ee

\noindent
in the presence of a gauge function $B(q_i, t)$, where $X^{(1)}$ is the first prolongation of the vector field $X$ given by,
\be X^{(1)} = X + \sum_{i}\Big[(\dot\alpha_{i} - \dot \eta\dot q_{i})\frac{\partial}{\partial\dot q_{i}}\Big]\;\;\&\;\;  X = \eta\frac{\partial}{\partial t} +\sum_{i} \alpha_i\frac{\partial}{\partial  q_{i}} ,\ee

\noindent
with, $\alpha_i = \alpha_i(q_i, t), ~\eta = \eta(q_i, t)$, then there exists a conserved current,
\be I = \sum_{i}(\alpha_i - \eta \dot q_{i} )\frac{\partial L(q_{i}, \dot q_{i}, t)}{\partial \dot q_{i}} + \eta L(q_{i}, \dot q_{i}, t) - B(q_{i}, t) = \sum_{i}\alpha_i p_{i} - B(q_i, t) - \eta(q_i, t)H(q_i, p_i, t).\ee

\noindent
In some earlier works \cite{i, i2} it has been shown that under infinitesimal co-ordinate and temporal transformations ($q_i' = q_i + \epsilon\alpha_i(q_i,t)$ and $t' = t + \epsilon\eta(q_i, t)$) time dependence may be introduced in a Lagrangian which does not contain time explicitly, through a time dependent gauge function. Invariance of Hamilton's principal function finally yields,

\be\begin{split} \int L(q_{i}, \dot q_{i}, t) dt & = \int L(q_{i}, \dot q_{i}) dt - \epsilon\frac{d}{dt}\int\left[\eta\left(L(q_i,\dot q_i, t) - \sum\dot q_i\frac{\partial L(q_i, \dot q_i, t)}{\partial \dot q_i}\right) + \sum\alpha_{i} \frac{\partial L(q_i, \dot q_i, t)}{\partial \dot q_i} - B(q_i, t)\right]dt\\
&=\int L(q_{i}, \dot q_{i}) dt - \epsilon\frac{d}{dt}\int\left[-\eta H + \sum\alpha_{i} p_i - B(q_i, t)\right]dt,\end{split}\ee

\noindent
retaining only up to first order term in Taylor expansion. It is apparent that the conserved current (3) may be obtained from (4) only if the Lagrangian in the left hand side and that in the right hand side of the above expression (4) get cancelled. This is possible provided the Lagrangian appearing in the left hand side is time independent. Remember that time dependence has been generated in the Lagrangian only through a time dependent gauge function $B(q_i, t)$ \cite{i}. Thus for a time independent Lagrangian (as in the case of gravity under consideration), the gauge function has to be time-independent as in the case of harmonic oscillator. Further, it is known that Noether integral is the Hamiltonian for the trivial Noether point symmetry $\frac{\partial}{\partial t}$. Other way round, the Hamiltonian is the generator of time translation and so conservation of the Hamiltonian requires $\eta$ to be constant. Further, in gravity, the Hamiltonian is not only conserved, but it is constrained to vanish. As a result, $\eta$ does not play any significant roll in Noether symmetry as is clearly observed from equation (4) and may even be set equal to zero. Time independence of the Lagrangian, the gauge term and $\eta$ automatically enforces time independence of $\alpha_i$. In the process, the integral of motion (3) reduces simply to

\be  I = \sum_{i}\alpha_i(q_i) p_{i} - B(q_i).\ee
Essentially if gauge turns out to be zero while solving Noether equations, the integral of motion remains the same as for Noether symmetry without gauge and no new result is expected. Although, it is clear that time translation is unnecessary, still we keep it explicitly in the following, to keep track with the earlier work \cite{j} performed in this context.

\noindent
Now, considering Robertson-Walker line element

\be ds^2 = -dt^2+a^2\left[\frac{dr^2}{1-kr^2} + r^2 d\theta^2+r^2\sin^2\theta d\varphi^2\right],\ee
the following Born-Infeld effective 4-dimensional action for a rolling tachyon field $\phi$ with Lagrangian density ${\mathcal{L(\phi)}} = - V(\phi)\sqrt{1 + \zeta\phi_{,\mu}\phi^{,\mu}}$ being minimally coupled to $F(R)$

\be A = \int  d^4 x\sqrt{-g}\left[F(R)-V(\phi)\sqrt{1 + \zeta\phi_{,\mu}\phi^{,\mu}}\right],\ee
\noindent
leads to a point Lagrangian

\be L = 6a \dot a^2 F' + 6a^2 \dot a\dot R F'' + a^3(F' R-F) + a^3 V(\phi)\sqrt{1 - \zeta\dot\phi^2},\ee
\noindent
for the spatially flat ($k = 0$) case, treating

\be R = 6\left(\frac{\ddot a}{a}+\frac{\dot a^2}{a^2}\right),\ee
as a constraint of the theory and effectively spanning the Lagrangian by a set of configuration space variables $(a,R,\phi,\dot a,\dot R,\dot  \phi,)$. In the above action $\zeta$ is treated as the coupling constant required to make the kinetic part of the action dimensionless. For the above point Lagrangian the Noether equation (1) reads,

\be\begin{split}& \alpha\Big[6\dot a^2 F'+12a\dot a\dot R F''+3a^2(F'R - F) + 3a^2 V\sqrt{1-\zeta\dot\phi^2}\Big]+\beta\Big[6a\dot a^2F''+6a^2\dot a\dot RF'''+a^3RF''\Big]+\gamma\Big[a^3V_{,\phi}\sqrt{1-\zeta\dot\phi^2}\Big]\\
&+\Big[(\alpha_{,t}-\dot a\eta_{,t})+\alpha_{,a}\dot a + \alpha'\dot R + \alpha_{,\phi}\dot\phi -\eta_{,a}\dot a^2 -\eta'\dot a\dot R -\eta_{,\phi}\dot a\dot\phi\Big][12 a\dot aF'+6a^2\dot R F'']\\
&+\Big[(\beta_{,t}-\dot R\eta_{,t})+\beta_{,a}\dot a + \beta'\dot R + \beta_{,\phi}\dot\phi -\eta_{,a}\dot a\dot R -\eta'\dot R^2 -\eta_{,\phi}\dot R\dot\phi\Big][6 a^2\dot aF'']\\
&+\Big[(\gamma_{,t}-\dot \phi\eta_{,t})+\gamma_{,a}\dot a + \gamma'\dot R + \gamma_{,\phi}\dot\phi -\eta_{,a}\dot a\dot\phi -\eta'\dot R\dot\phi -\eta_{,\phi}\dot\phi^2\Big]\left[-\frac{\zeta a^3 V\dot\phi}{\sqrt{1-\zeta\dot\phi^2}}\right]\\
&+\Big[\eta_{,t}+\eta_{,a}\dot a+\eta'\dot R+\eta_{,\phi}\dot\phi\Big]\Big[ 6a \dot a^2 F' + 6a^2 \dot a\dot R F'' + a^3(F' R-F) + a^3 V\sqrt{1 - \zeta\dot\phi^2}\Big]= B_{,a}\dot a+B'\dot R + B_{,\phi}\dot\phi+B_{,t},\end{split}\ee

\noindent
where, dash $(')$ represents derivative with respect to $R$. Note that we have kept both time translation and a time dependent gauge function to track with the work performed by Jamil et al \cite{j}. Now, equating coefficients as usual, we obtain a large over-determinant set of Noether equations for $F''(R) \ne 0$. These are

\begin{eqnarray}
   \eta_{,a} =\eta' = \eta_{,\phi} &=& 0 \\
   \alpha' = \alpha_{,\phi} &=& 0 \\
   \beta_{,\phi} &=& 0  \\
   \gamma_{,t} = \gamma' = \gamma_{,a} &=& 0 \\
   B_{,\phi} &=& 0 \\
   \gamma_{,\phi} - \eta_{,t} &=&0\\
   6a^2 F''\alpha_{,t} &=& B'\\
   12 aF'\alpha_{,t} + 6a^2 F''\beta_{,t}&=& B_{,a} \\
   a^2(3\alpha+a\eta_{,t})(F' R - F)+a^3\beta R F''  &=& B_{,t}\\
   \alpha F' + a\beta F''+2a\alpha_{,a} F'+a^2 \beta_{,a} F'' - aF'\eta_{,t} &=& 0 \\
   2a\alpha F'' + a^2 \beta F'''+a^2 F''(\alpha_{,a}+\beta'-\eta_{,t}) &=& 0  \\
   3\alpha V +a\gamma V_{,\phi} +a V \eta_{,t} &=&0
     \end{eqnarray}
Before we proceed let us mention that if $V(\phi) = 0$, the action (7) corresponds to pure $F(R)$ theory of gravity, for which nothing other than $F(R) \propto R^{\frac{3}{2}}$ is possible \cite{c}. In what follows, we shall assume $V(\phi) \ne 0$ in equation (22) and so it will never be possible to recover pure $F(R)$ case. Now, in view of equations (11) through (15), it is apparent that $\eta = \eta(t)$, $\alpha = \alpha(t,a)$, $\beta = \beta(t, a, R)$, $\gamma = \gamma(\phi)$ and $B = B(t, a, R)$. Therefore equation (16) dictates that $\eta$ may be at most linear in $t$. Thus equation (22) clearly states that $\alpha$ has to be time independent (as $\gamma$ is) and it should be linear in the scale factor $a$. Thus taking $\alpha = c_4 a$, where $c_4$ is a constant, it is clear that $\beta$ has to be independent of $t$ and $a$ in view of equations (20) and (21) and the gauge term $B$ must be independent of $R$ as is apparent from equation (17). Further, since both $\alpha$ and $\beta$ are time independent, therefore the gauge term further becomes independent of the scale factor $a$ in view of equation (18). Finally, equation (19) is satisfied provided $B \ne B(t)$. In the process the gauge term $B$ turns out to be a constant and hence plays no role in Noether symmetry. Thus one can set $B = 0$, without loss of generality. In view of the above analysis, we now have,

\be \eta = c_1 t + c_3,\;\; \alpha = c_4 a,\;\;\beta = \beta(R),\;\;\gamma = c_1\phi + c_2\;\;and\;\;B = 0,\ee
where, $c_1, c_2, c_3$ and $c_4$ are constants. Having obtained explicit forms of $\eta,\; \alpha$ and $\gamma$, we are now left to find the explicit forms of $\beta(R)$, $F(R)$ and the potential $V(\phi)$ from the set of over-determinant equations (19) through (22) which in view of (23) now get reduced to following four equations,
 \begin{eqnarray}
   (3c_4 + c_1)(F' R - F)+ F'' R\beta &=& 0\\
   (3c_4 - c_1)F' + \beta F''&=& 0 \\
   \beta F'''+ F''(3c_4 + \beta'- c_1)&=& 0\\
   (3c_4 + c_1)V + (c_1\phi + c_2)V_{,\phi}  &=&0
 \end{eqnarray}

\subsection{Reviewing Jamil et al's work}
Equation (27) may now be solved for the following form of the potential

\be V = V_{10}(c_1\phi + c_2)^{-\left(\frac{3c_4+c_1}{c_1}\right)} = c_1V_{10}\left(\phi + \frac{c_2}{c_1}\right)^{-\left(\frac{3c_4+c_1}{c_1}\right)},\ee
$V_{10}$ being a constant. However, the authors \cite{j} claimed the following form of the potential

\be V = V_0(\phi + \phi_0)^{-4},\ee
which is possible only under the choices, $V_0 = c_1 V_{10}$, $\phi_0 = c_2/c_1$ and in particular, $c_1 = c_4$, which as we shall show shortly, create severe problem. One can easily check that under the choice $c_1 = c_4$, equation (26) upon integration yields equation (25), provided the constant of integration is zero. Thus we are now left with two independent equations (24) and (25) to find the form of $F(R)$ and $\beta(R)$. However, equation (25) now reads

\be \beta F'' = -2c_1 F'.\ee
Now eliminating $\beta$ between equations (24) and (30) one ends up with

\be F(R) \propto R^2\ee
in a straightforward manner and $\beta$ gets solved as $\beta = -2c_1 R$. At this end the authors \cite{j} obtained two conserved current $I_1$ and $I_2$. $I_1$ is the invariance under time translation, as stated correctly by the authors (here we point out a typographical error in equation (31) of \cite{j} . $I_1$ should be $I_1 = \tau\Big[L - \big(\dot a\frac{\partial L}{\partial\dot a} + \dot R\frac{\partial L}{\partial\dot R}+\dot \phi\frac{\partial L}{\partial\dot\phi}\big)\Big]$ instead, where $\tau$ used in \cite{j} stands for $\eta$ here). Nevertheless, the third bracketed term is the Hamiltonian, which is constrained to vanish. Thus if $I_1 = 0$ is substituted in $I_2$ (as already mentioned in section 2), the conserved current is simply,

\be I_2 = \sum \alpha_i p_i =  \left[12 F_0 \dot R + \frac{\zeta V_0\dot\phi}{(\phi + \phi_0)^3\sqrt{1-\zeta\dot\phi^2}}\right]a^3.\ee
Conserved current is not an independent equation, rather it is the first integral of certain combination of the field equations. Thus it is essential to check if the above conserved current obtained for $F(R) = F_0 R^2$ and $V = V_0(\phi + \phi_0)^{-4}$, satisfies the field equations, which was not performed by the authors \cite{j}. Now the corresponding field equations are

\be F_0\left[2\frac{\ddot a}{a} + \frac{\dot a^2}{a^2} + \frac{\ddot R}{R}+ 2 \frac{\dot a\dot R}{a R} - \frac{R}{4}\right] + \frac{V}{4 R}\sqrt{1-\zeta\dot\phi^2} =0.\ee

\be \ddot\phi + 3\frac{\dot a}{a}\dot\phi -3\zeta \frac{\dot a}{a}\dot\phi^3+ \frac{V_{,\phi}}{\zeta V}(1 - \zeta \dot\phi^2) = 0.\ee

\be H = F_0[12a \dot a^2 R + 12 a^2 \dot a\dot R -a^3 R^2] + \frac{a^3V(\phi)}{\sqrt{1-\zeta\dot\phi^2}}=0.\ee
It is not difficult to check that the above conserved current ($I_2$) satisfies the field equations only under the trivial condition $R = 0$ together with the condition $V_0 = 0$, i.e., for vanishing potential, which leads to inconsistency, since it has been restricted at the beginning. Thus $F(R) \propto R^2$ is not a symmetry of the action under consideration and the work performed by Jamil et al \cite{j} is completely wrong. One can even look at the consequence in a straight-forward manner. It is well-known that for Noether point symmetry $\frac{\partial}{\partial t}$, the Noether integral is the Hamiltonian, i.e., $I_1$ is the Hamiltonian $H$. But here, $I_1 = \eta H = (c_1 t + c_3)H$. Thus unless $\eta = 1$, implying $c_1 = 0$ and $c_3 = 1$, Hamiltonian is not obtained as the integral of motion. However, since $c_1 = c_4$, it can not be set equal to zero at this stage, as it makes $\alpha = \beta = 0$ and $\gamma =$ constant, while the form of the potential (28) becomes undefined. Thus the Hamiltonian is never recovered as Noether integral. Therefore, to recover the Hamiltonian constraint equation as an outcome of Noether point symmetry $\frac{\partial}{\partial t}$, one should start with $c_1 = 0$ a-priori, which we consider in the following subsection.

\subsection{No need to consider time translation, so, $\eta = c_3 =1$.}
\noindent
It must have been clear by this time why earlier authors did not consider time translation. Likewise, if one starts with $\eta = 1$, which implies $c_1 = 0$ and $c_3 =1$, $\gamma$ turns out to be a constant and may be set as $\gamma = c_2 = 1$, without loss of generality. thus equations (24) through (27) take the following forms,

\begin{eqnarray}
   3 c_4 (F' R - F)+ R F'' \beta & = & 0\\
   3 c_4 F' + F''\beta & = & 0 \\
   \beta F''' + F''(3 c_4 + \beta') & = & 0  \\
   3 c_4 V + V_{,\phi} & = & 0
\end{eqnarray}
Now, just multiplying equation (37) by $R$ and comparing it with equation (36), one can observe that $F(R) = 0$. Thus search for a form of $F(R)$ by imposing Noether symmetry in the action (7) went in vain.

\section{In search of Noether symmetry for a more general tachyonic action}

Having obtain null result in connection with Noether symmetry for Born-Infeld action being coupled to $F(R)$, let us now turn our attention to a more general action containing a linear curvature invariant term being non-minimally coupled to Born-Infeld-$F(R)$ action (7). However, as we have already mentioned that the gravitational Hamiltonian is not only conserved but is constrained to vanish and therefore is a part and parcel of the Einstein's equation, viz., the $(^0_0)$ component, so time translation is overall unnecessary. Further, in all our earlier analysis we have shown that a gauge term does not contribute to the Noether equations, since it becomes constant and therefore may be set equal to zero. Therefore, in the following analysis neither do we consider time translation nor a gauge term. In the Robertson-Walker line element (6) the following action
\be A = \int  d^4 x\sqrt{-g}\left[h(\phi)R + {\mathcal B}F(R) - V(\phi)\sqrt{1 + \zeta\phi_{,\mu}\phi^{,\mu}}\right],\ee
\noindent
leads to the point Lagrangian
\be L =6a \dot a^2 h+ 6a^2 \dot a\dot\phi h_{,\phi}-6kah+ {\mathcal B}[6a \dot a^2 F' + 6a^2 \dot a\dot R F'' + a^3(F-F' R)-6kaF'] +a^3 V(\phi)\sqrt{1 - \zeta\dot\phi^2},\ee
\noindent
following the same technique as before. For the above point Lagrangian the Noether equation (1) reads,
\[\alpha\Big[6\dot a^2h+ 12a\dot a\dot\phi h_{,\phi}-6kh + {\mathcal B}[6 \dot a^2 F' +12a\dot a\dot R F''+3a^2(F-F'R)-6kF'] + 3a^2 V\sqrt{1-\zeta\dot\phi^2}\Big]\]
\[+{\mathcal B}\beta\Big[6a\dot a^2F''+6a^2\dot a\dot RF'''-a^3RF''-6kaF''\Big]+\gamma\Big[6a\dot a^2h_{,\phi}+6a^2\dot a\dot\phi h_{,\phi\phi}-6kah_{,\phi}+\ a^3V_{,\phi}\sqrt{1-\zeta\dot\phi^2}\Big]\]
\[+\Big[\alpha_{,a}\dot a + \alpha'\dot R + \alpha_{,\phi}\dot\phi \Big]\Big[12a\dot ah+6a^2h_{,\phi}\dot\phi+{\mathcal B}[12 a\dot aF'+6a^2\dot R F'']\Big]
+\Big[\beta_{,a}\dot a + \beta'\dot R + \beta_{,\phi}\dot\phi \Big][6 {\mathcal B}a^2\dot aF'']\]
\be+\Big[\gamma_{,a}\dot a + \gamma'\dot R + \gamma_{,\phi}\dot\phi \Big]\left[6a^2\dot ah_{,\phi}-\frac{ \zeta a^3 V\dot\phi}{\sqrt{1-\zeta\dot\phi^2}}\right]=0,\ee

\noindent
Now, equating coefficients as usual, we obtain the following over-determinant set of Noether equations
\begin{eqnarray}
   \alpha' = \alpha_{,\phi} &=& 0 \\
 2\alpha h_{,\phi}+a\gamma h_{,\phi\phi}+a h_{,\phi}\alpha_{,a} + a h_{,\phi}\gamma_{,\phi} + {\mathcal B} a F''\beta_{,\phi} &=& 0  \\
  \gamma' = \gamma_{,a} &=& 0 \\
   \gamma_{,\phi}  &=&0\\
   {\mathcal B}[3a^2\alpha(F-F' R)-a^3\beta R F''-6\alpha k F'-6\beta k a F'']-6\alpha k h  &=& 0\\
  \alpha h+a\gamma h_{,\phi}+2ah\alpha_{,a} + {\mathcal B}[\alpha F' + a\beta F''+2a\alpha_{,a} F'+a^2 \beta_{,a} F''] &=& 0 \\
   2a\alpha F'' + a^2 \beta F'''+a^2 F''(\alpha_{,a}+\beta') &=& 0  \\
   3\alpha V +a\gamma V_{,\phi}  &=&0
     \end{eqnarray}

\subsection{Solutions}
The above set of equations (43) through (50) imply $\alpha = \alpha(a)$, $\beta = \beta( a, R, \phi)$, $\gamma = \gamma_0$, where $\gamma_0$ is a constant. Equation (44) then determines $\alpha$ as a linear function of the scale factor, viz.,
\be \alpha = \alpha_0 a\ee
and equation (44) gets reduced to
\be 3\alpha_0 h_{,\phi} + \gamma_{0}h_{,\phi\phi} + {\mathcal B}F''\beta_{,\phi} = 0.\ee
Equation (47) is then satisfied only for spatially flat $k = 0$ case and thus reduces to
\be 3\alpha_0(F - F'R) = \beta R F''\ee
forcing $\beta$ to be a function of $R$ only, ie., $\beta = \beta(R)$. Equation (48) then in view of equation (47) makes $F(R)$ a linear function of $R$ as $F(R) = F_0 R$, $F_{0}$ being a constant. Thus search for Noether symmetry for $F''(R) \ne 0$ again went in vain. Nevertheless, the coupling parameter $h(\phi)$ gets solved as,
\be h(\phi) = F_0 e^{-\frac{\phi}{\phi_{0}}}\ee
where, $\phi_0 = \frac{\gamma_0}{3\alpha_0}$ is a constant. Equation (49) is then trivially satisfied while equation (50) yields the following form of the potential,
\be V(\phi) = V_0 e^{-\frac{\phi}{\phi_0}}.\ee
The above forms of the potential $V(\phi)$ and the coupling parameter $h(\phi)$ along with the conserved current had been found earlier by de Souza and Kremer \cite{t8} as a consequence of Noether symmetry for linear gravity and further, they had explicitly studied the cosmological evolution. Hence, we leave our discussion here. Nonetheless, we observe that $F(R)$ theory of gravity does not admit Noether symmetry even for a general Born-Infeld action and so our earlier conclusion that `it is not possible to find a form $F(R)$ other than the trivial and very special one ($R^{\frac{3}{2}}$) by imposing Noether symmetry' - stands.

\section{On the absence of Noether symmetry and exploring the special feature of $R^{\frac{3}{2}}$}

We have mentioned that starting from $F(R) = R^2$ a-priori, Noether symmetry is obtainable for a scalar-tensor theory of gravity \cite{e}. Further it has been proved that $F(R) = R^2$ leads to a scale invariant action \cite{sd}. So, it is expected that starting from arbitrary $F(R)$ if Noether symmetry is claimed, it should end up with $F(R) \propto R^2$. But we have not been able to recover this result. This contradiction puts up doubt in treating $R$ as an auxiliary variable for canonical formulation of $F(R)$ theory of gravity, since an auxiliary variable $Q$ different from $R$ has been considered for canonical formulation of $R^2$ theory of gravity. Let us briefly describe the issue.\\

In the quantum domain observable depends on the choice of momentum, while momentum is different for different choice of auxiliary variable. The canonical formulation of $R^2$ gravity in view of the auxiliary variable $Q = \frac{\partial A}{\partial \ddot h_{ij}}$ ($A$ being the action) leads to a Schr{\"o}dinger like quantum dynamics, with a hermitian effective Hamiltonian leading to the straightforward probability interpretation \cite{f1, f2, f3, f4, f5, f6}, provided the total derivative terms in the action are taken care of a-priori. From metric variation principle it is known that $R^2$ theory of gravity must be supplemented by a boundary term $\Sigma = 4\beta \int ~(^4R) K\sqrt {h} ~d^3 x$, where symbols have their usual meaning. It was shown \cite{f5, f6} that to obtain Schr{\"o}dinger like quantum equation as mentioned, it is required first to express the action in terms of the first fundamental form $h_{ij}$ and then to split the above boundary term into $\Sigma = \sigma_1 + \sigma_2$ where, $\sigma_1 = 4\beta \int ~(^3R) K\sqrt {h} ~d^3 x$ and $\sigma_2 = 4\beta\int ~(^4R-^3R) K\sqrt {h} ~d^3 x$. Canonical programme then follows by eliminating the available total derivative term from the action, which gets cancelled with the boundary term $\sigma_1$, and then introducing the auxiliary variable, $Q = \frac{\partial A}{\partial \ddot h_{ij}}$, as suggested by Horowitz \cite{g} thereafter. Thus $Q$ is different from $R$ in the $R^2$ theory of gravity. It is not possible to follow such technique for an action containing a general $F(R)$ theory of gravity. It is also important to mention that classical field equations require derivative of momentum and so arbitrary choice of auxiliary variable reproduces correct classical field equations. However, as for quantization one requires momentum (rather than its derivative, which is the reason for obtaining different quantum dynamics with different auxiliary variables) likewise, for Noether symmetry one again requires momentum instead of its derivative and this may be the reason why it could not reproduce Noether symmetry for $F(R) = R^2$ theory of gravity already available in the literature \cite{e}. All the beauty of $F(R)$ theory of gravity observed in the context of cosmological data fitting are outcome of scalar-tensor equivalence, since otherwise it is almost impossible to find exact solutions. Scalar-tensor equivalence is a mathematical artifact and it gives totally different quantum description \cite{qd} in comparison to one discussed earlier \cite{f4, f5}. The difference at the classical level has also been established to some extent \cite{ineq1}, \cite{ineq2}. Thus the prospect of $F(R)$ theory of gravity appears to be at stake, unless one can find a better way out to handle the theory. Indeed a better technique has been expatiated earlier by finding a general (non-Noether) conserved current for $F(R)$ theory of gravity being minimally coupled to scalar-tensor theory of gravity \cite{k}. The technique was also found useful to extract exact solutions of the theory \cite{d, l}.\\

The obvious question is - how then $F(R) \propto R^{\frac{3}{2}}$ had been found in vacuum and in radiation dominated era? Let us brief the beauty of $F(R)\propto R^{\frac{3}{2}}$ in isotropic space-time \cite{c}. Remember that in metric variation technique, the field equations are obtained by varying the action with respect to $g_{\mu\nu}$, while the scale factor ($a = \sqrt{h_{ij}}$) is taken as the basic variable to express the canonical action. Instead, if we take $h_{ij} = a^2 = z$ to be the basic variable then $R = 6(\frac{\ddot z}{2z}+\frac{k}{z})$ and so the action
\be A = \int \sqrt{-g}d^4 x F(R) + 2 \int_{\Sigma}\sqrt{h}F_{,R}K d^3 x \ee
for $F(R) = R^{\frac{3}{2}}$ reads

\be A = 3\sqrt 3  \int(\ddot z + 2k)^{\frac{3}{2}}dt - 2\int_{\sigma}\frac{3}{2}z^{\frac{3}{2}}\sqrt R K d^3 x\ee
Introducing an auxiliary variable

\be Q = \frac{\partial A}{\partial \ddot z} =\frac{9\sqrt 3}{2}(\ddot z + 2k)^{\frac{1}{2}}\ee
(which is clearly different from $R$) in the action, and removing appropriate surface terms, the canonical form of the action is
\be A = \int \left[-\dot Q\dot z + 2kQ - \frac{4 Q^3}{729 B^2}\right]dt\ee
Clearly $z$ is cyclic and a Noether conserved current
\be \frac{d}{dt}(a\sqrt R) = {\mathrm {constant}}\ee
is apparent, which may be solved trivially to yield
\be a = [a_4 t^4 + a_3 t^3 + a_2 t^2 + a_1 t + a_0]^{\frac{1}{2}}\ee
while $Q$ variation equation only reproduces the definition of $Q$ given above. Thus $F(R) = R^{\frac{3}{2}}$ leads to a trivial Noether current when viewed in terms of the basic variable $h_{ij}$. The above solution obtained by several authors clearly leads to power law inflation. It is also admissible in the matter dominated era, leading to present acceleration, but early deceleration remains absent.

\section{Handling $F(R)$ theory of gravity in view of a general conserved current}

In the absence of Noether symmetry of $F(R)$ theory of gravity coupled to tachyon field, even if a suitable form of $F(R)$ is chosen by hand, it is extremely difficult, if not impossible to find exact solution. However, in the literature there exists a technique to find a conserved current for nonminimally coupled scalar-tensor theory of gravity \cite{k1} and also for $F(R)$ being minimally coupled to a non-minimal scalar-tensor theory of gravity \cite{d, k, l}. This conserved current has been found useful to extract solutions. Here, we explore the same corresponding to the action (40). We first briefly review the issue of conserved current already explored \cite{d, k, l}. The field equations corresponding to the following action containing scalar tensor theory of gravity in the presence of $F(R)$
\be
A = \int\left[h(\phi) ~R + B F(R) -\frac{\omega(\phi)}{\phi}\phi,_{\mu}\phi^{,\mu} -V(\phi)-\kappa L_{m}\right]\sqrt{-g}~d^4x, \ee
are
\be\begin{split}
&h\left(R_{\mu\nu}- \frac{1}{2}g_{\mu\nu}R\right)+h^{;\alpha}_{;\alpha}g_{\mu\nu}-h_{;\mu
;\nu} -\frac{\omega}{\phi}\phi_{,\mu}\phi_{,\nu}+
\frac{1}{2}g_{\mu\nu}\left(\frac{\omega}{\phi}\phi_{,\alpha}\phi^{,\alpha}+V(\phi)\right)\\
& +B\left[F'R_{\mu\nu}-\frac{1}{2}F g_{\mu\nu}+
{{(F')}^{;\alpha}}_{;\alpha}~ g_{\mu\nu}-(F')_{;\mu;\nu}\right]=\frac{\kappa}{2}T_{\mu\nu} \end{split}\ee

\be
R h_{,\phi}+2\frac{\omega}{\phi}\phi^{;\mu}_{;\mu}+\left(\frac{\omega_{,\phi}}{\phi}
-\frac{\omega}{\phi^2}\right)\phi^{,\mu}\phi_{,\mu}-V_{,\phi}(\phi) = 0\ee

The trace of equation (63) is the following,

\be
R h-3h^{;\mu}_{;\mu}-\frac{\omega}{\phi}\phi^{,\mu}\phi_{,\mu}-2V
-B[R F' +3\Box(F')-2F] = -\frac{\kappa}{2}T^{\mu}_{\mu}. \ee
Now eliminating the first term between equations (64) and (65), then substituting $\Box h = h_{,\phi}\phi^{,\mu}_{~;\mu} + h_{,\phi\phi}\phi_{,\alpha}\phi^{,\alpha}$ and following a little algebra, one can arrive at the following equation

\be
\left[(3h_{,\phi}^2+2h\frac{\omega}{\phi})^{\frac{1}{2}}\phi^{;\mu}\right]_{;\mu}+
\left(3h_{,\phi}^2+2h\frac{\omega}{\phi}\right)^{-\frac{1}{2}}\left[B
h_{,\phi}[R F' +3\Box(F')-2F] -\frac{\kappa}{2}h_{,\phi}~T^{\mu}_{\mu}
-h^3\left(\frac{V}{h^2}\right)_{,\phi}\right]=0. \ee In view of which one can conclude that under the following condition,

\be B [R F' +3\Box (F')-2F] = \frac{\kappa}{2}~T^{\mu}_{\mu} + \frac{h^3}{h_{,\phi}}\left(\frac{V}{h^2}\right)_{,\phi} \ee
there exists a conserved current $J^\mu$, where

\be
J^{\mu}_{~;\mu}=\left[\left(3h_{,\phi}^2+2h\frac{\omega}{\phi}\right)^{\frac{1}{2}}\phi^{;\mu}\right]_{;\mu}
= 0. \ee
Further, assuming $V \propto h^2$, one finds that the condition (67) for the existence of the conserved current does not depend on the scalar field and the choice of potential. Thus, the cosmological evolution only depends on the form of $F(R)$, which may be chosen by hand. Note that $T^{\mu}_{\mu}$ vanishes in vacuum and radiation dominated era and so in these era the fluid distribution does not play any role. In fact if one takes (say for example) $F(R) \propto R^2$, power law inflation is realized in vacuum era. Inflation makes the space-time flat and so assuming $k = 0$ in the isotropic and homogeneous metric, the cosmological evolution of the scale factor in the radiation era, behaves like Friedmann solution $(a \propto \sqrt t)$. Thus higher order curvature invariant term does not affect Baryogenesis, Nucleosynthesis and Structure formation (see \cite{l}). Let us now turn our attention to find similar conserved current corresponding to the action (40) containing tachyonic field. The field equations corresponding to the action (40) containing a matter part in addition are the following

\be h\left(R_{\mu\nu} - \frac{1}{2}g_{\mu\nu}R\right) + \Box h g_{\mu\nu} - h_{;\mu;\nu} + {\mathcal B}\Big[F'R_{\mu\nu}
  -\frac{1}{2}F g_{\mu\nu} + {(\Box F')}~ g_{\mu\nu} -(F')_{;\mu ;\nu}\Big] = \frac{1}{2}\left({\mathcal T_{\mu\nu}} + T_{\mu\nu}\right).\ee

\be R h_{,\phi} + \left(\frac{\zeta V \phi^{,\alpha}}{\sqrt{1 + \zeta\phi_{,\mu}\phi^{,\mu}}}\right)_{,\alpha} - V_{,\phi}\sqrt{1 + \zeta\phi_{,\mu}\phi^{,\mu}} = 0, \ee
where, ${\mathcal T_{\mu\nu}}$ is the energy-momentum tensor for the Tachyon field given by

\be \mathcal{T_{\alpha\beta}} = g_{\alpha\beta} V(\phi)\sqrt{1 + \zeta\phi_{,\mu}\phi^{,\mu}}- \frac{\zeta V \phi_{\alpha}\phi_{\beta}}{\sqrt{1 + \zeta\phi_{,\mu}\phi^{,\mu}}}\ee
while $T_{\mu\nu}$ is that for usual the matter field. Trace of equation (69) is

\be  R h - 3\Box h - {\mathcal B} [R F' + 3 \Box F' - 2F] =  -\frac{1}{2}\left({\mathcal T_{\mu}^{\mu}} + T_{\mu}^{\mu}\right)   .\ee
Now eliminating the first terms between equations (70) and (72) we obtain

\be\begin{split}&\left(3h_{,\phi}^2 + \frac{\zeta V h}{\sqrt{1 + \zeta\phi_{,\mu}\phi^{,\mu}}}\right)\phi^{,\alpha}_{~;\alpha} + 3h_{,\phi}h_{,\phi\phi}\phi_{,\alpha}\phi^{,\alpha}+ \frac{\zeta h V_{,\phi}\phi_{,\alpha}\phi^{\alpha}}{\sqrt{1 + \zeta\phi_{,\mu}\phi^{,\mu}}} + \frac{\zeta V h \phi^{,\alpha}}{(\sqrt{1 + \zeta\phi_{,\mu}\phi^{,\mu}})_{,\alpha}} + \frac{\zeta V h_{,\phi}\phi_{,\alpha}\phi^{,\alpha}}{2\sqrt{1 + \zeta\phi_{,\mu}\phi^{,\mu}}}\\
&-2V h_{,\phi} \sqrt{1 + \zeta\phi_{,\mu}\phi^{,\mu}} - V_{,\phi}h\sqrt{1 + \zeta\phi_{,\mu}\phi^{,\mu}} + h_{,\phi}\left[{\mathcal{B}}(R F' + 3\Box F' -2F) - \frac{T^{\mu}_{\mu}}{2}\right] = 0\end{split}\ee
where, we have substituted $\Box h = h_{,\phi}\phi^{,\mu}_{~;\mu} + h_{,\phi\phi}\phi_{,\alpha}\phi^{,\alpha}$ and the trace $\mathcal{T^{\mu}_{\mu}}$ of the energy-momentum tensor corresponding to the Tachyon field. The above equation can be rearranged as,

\be\begin{split}&\left[\left(\sqrt {3h_{,\phi}^2 + \frac{\zeta V h}{\sqrt{1 + \zeta\phi_{,\mu}\phi^{,\mu}}}}\right)\phi^{,\alpha}\right]_{;\alpha} +  \left( {3h_{,\phi}^2 + \frac{\zeta V h}{\sqrt{1 + \zeta\phi_{,\mu}\phi^{,\mu}}}}\right)^{-\frac{1}{2}}\times \\
&\left[ h_{,\phi}\left(\mathcal{B}\Big(R F' + 3\Box F' -2F\Big) -\frac{T^{\mu}_{\mu}}{2}\right) + \frac{\zeta h}{2}\left(\frac{V}{\sqrt{1 + \zeta\phi_{,\mu}\phi^{,\mu}}}\right)_{;\alpha} - \sqrt{1 + \zeta\phi_{,\mu}\phi^{,\mu}} h^3 \left(\frac{V}{h^2}\right)_{,\phi}\right]=0 \end{split}.\ee
One can now conclude that there exists a conserved current

\be J^{\alpha}_{~;\alpha} = \left[\left(\sqrt {3h_{,\phi}^2 + \frac{\zeta V h}{\sqrt{1 + \zeta\phi_{,\mu}\phi^{,\mu}}}}\right)\phi^{,\alpha}\right]_{;\alpha} = \frac{1}{\sqrt{-g}}\left[\left(\sqrt {3h_{,\phi}^2 + \frac{\zeta V h}{\sqrt{1 + \zeta\phi_{,\mu}\phi^{,\mu}}}}\right)\phi^{,\alpha}\sqrt{-g}\right]_{,\alpha} = 0\ee
provided

\be\left[ h_{,\phi}\left(\mathcal{B}\Big(R F' + 3\Box F' -2F\Big) -\frac{T^{\mu}_{\mu}}{2}\right) + \frac{\zeta h}{2}\left(\frac{V}{\sqrt{1 + \zeta\phi_{,\mu}\phi^{,\mu}}}\right)_{;\alpha} - \sqrt{1 + \zeta\phi_{,\mu}\phi^{,\mu}} h^3 \left(\frac{V}{h^2}\right)_{,\phi}\right] = 0.\ee
Note that treating the scalar curvature $R$ as a single variable, we had only a couple of equations, (69) and (70) to solve $R, F(R), \phi, h(\phi) ~\mathrm {and}~ V(\phi)$. So for exact solution we require altogether three physically reasonable assumptions. Above conserved current has been constructed out of the two field equations. So taking one of the field equations (say 70), the conserved current (75) and the condition (76) altogether gives us three equations. We therefore need two more assumptions to extract exact solution for the system. One of these may be set by separating the condition (76), so that cosmic evolution, ie., the solution of $R$ becomes field independent and depends only on the form of $F(R)$ chosen by hand, as before. Thus, we separate the condition (76) for the existence of the conserved current as,

\be \mathcal{B}\Big(R F' + 3\Box F' -2F\Big) = \frac{T^{\mu}_{\mu}}{2}\ee
\be \frac{\zeta h}{2}\left(\frac{V}{\sqrt{1 + \zeta\phi_{,\mu}\phi^{,\mu}}}\right)_{;\alpha} - \sqrt{1 + \zeta\phi_{,\mu}\phi^{,\mu}} h^3 \left(\frac{V}{h^2}\right)_{,\phi} = 0\ee
since $h_{,\phi} \ne 0$. Clearly the condition (67) (under the choice $V \propto h^2$) is the same as obtained here in (77). Therefore, the cosmological evolution is independent of choice of the field, even if it is tachyon. Thus we have completed our goal to find a general conserved current for tachyonic field such that, the cosmological evolution does not depend on the field and the choice of potential at all. Finally, as soon as a form of $F(R)$ would be chosen by hand one can, in principle, have explicit solution of the field equations under consideration. It is important to note that the cosmological evolution is dictated only by the form of $F(R)$ and it does not in any way depend on the linear curvature invariant term. In the following we shall take up Robertson-Walker line element (6) to explicitly demonstrate the applicability of the treatment developed. Equation (74) states that there exists a conserved current, viz.,

\be a^3\dot\phi \sqrt{3 h_{,\phi}^2 + \frac{\zeta V h}{\sqrt{1 - \zeta\dot\phi^2}}} = C\ee
$C$ being a constant, provided

\be   \zeta\dot\phi\ddot \phi + (1 - \zeta\dot\phi^2)\Big[\dot\phi - \frac{2}{\zeta}(1 - \zeta\dot\phi^2)\Big]\frac{V_{,\phi}}{V}  + \frac{4}{\zeta}(1 - \zeta\dot\phi^2)^2\frac{h_{,\phi}}{h}  = 0\ee
and

\be {\mathcal B}[R F' + 3 \Box F' - 2F] = \frac{1}{2} T_{\mu}^{\mu}.\ee
Note that if matter part ($T_{\mu\nu}$) is ignored for the time being, still we have a couple of independent field equations to solve for $F(R), h(\phi), V(\phi), \phi(t)$ and the scale factor $a(t)$. Thus it is required to make at least three physically reasonable assumptions to obtain explicit solutions of the system under consideration. Here, to obtain the conserved current, we have essentially made two assumptions viz., (80) and (81). Therefore, if a form of $F(R)$ is assumed, one can solve (81) for the scale factor $a(t)$ and the rest three viz., $h(\phi), V(\phi)$ and $\phi(t)$ may be solved in view of the conserved current (79), equation (80) and one of the field equations (say, 70). Interestingly enough, in the process curvature part (81) is decoupled from the field under consideration and the solution of the scale factor which emerges from equation (81) ie., the cosmic evolution remains independent of the form of field chosen. For example, in the flat case $k = 0$, it has been shown \cite{l} that for $F(R) = R^2$ equation (81) leads to power law inflationary solution in the vacuum dominated era, $a \propto \sqrt t$ in the radiation dominated era ($T^{\mu}_{\nu} = 0$), which is the result of the standard Friedmann model, despite strong coupling with Tachyon field has been incorporated and $a \propto t^{\frac{4}{3}}$ in the matter dominated era ($p = 0$). On the other hand, if $F(R) = R^{\frac{3}{2}}$ is chosen \cite{d}, equation (81) leads to $a \propto \sqrt t$ in the radiation dominated era and $a \propto t$ emerges as a particular solution in matter dominated era .

\section{Weak energy Limit}

Indeed, it is true that solar system puts up severe constraints on alternative theories of gravity \cite{st1, st2}. For an action
\[A = \int\sqrt{-g} d^4 x R^{n}\] the gravitational potential \cite{gpot} in the weak field limit \cite{weak} is found as

\[\Phi(r) = -\frac{Gm}{2r}\left[1+\left(\frac{r}{r_c}\right)^\beta\right]\]
where, $r_c$ is an arbitrary parameter varying within the range $(1 - 10^4)$AU taking into account the velocity of the earth to be $30 ~\mathrm{Km~ s^{-1}}$ \cite{st1} while $\beta$ is related to $n$ as
\[\beta = \frac{12n^2-7n-1-\sqrt{36n^4+12n^3-83n^2+50n+1}}{6n^2-4n+2}\]
Clearly, for $n = 1$, $\beta = 0$ and Newtonian gravitational field is recovered. For $n = \frac{3}{2}$, $\beta \sim 0.5$ Newtonian limit is not realized and such value of $\beta$ is ruled out from light bending data in the sun limb and planetary periods \cite{st1}.\\
Here, we provide additional constraint required to fit Newtonian limit for the more general action (40) being supplemented by ordinary matter action. In weak field approximation $g_{\mu\nu}=\eta_{\mu\nu} +h_{\mu\nu}$, where $\lvert h_{\mu\nu}\rvert \ll1$. Retaining only linear terms in $h_{\mu\nu}$ we have

\be R_{\mu\nu}\simeq \frac{1}{2}\Box h_{\mu\nu}~~ \mathrm {and}~~ R \simeq
\frac{1}{2} \Box h,~~ \mathrm {where}~~ h = h^{\mu}_{~\mu}\ee
along with the time-time component of field equation (63) taking $F(R) = R^{\frac{3}{2}}$ as

\begin{equation} \big(R_{00}-\frac{1}{2} g_{00}R\big)-{\mathcal{B}} R^{1/2}(3R_{00}-R g_{00}) -
\frac{3 {\mathcal B}}{2}R^{-\frac{3}{2}}\Big[\big(R\Box R -\frac{1}{2}R_{;\lambda}R^{;\lambda}\big)g_{00}
+ R R_{;0;0}-\frac{1}{2}R_{;0}R_{;0} \Big] =  {\mathcal T_{00}}+T_{00},
\end{equation}
provided $\phi \rightarrow$ constant with the expansion, so that $h(\phi) \rightarrow \frac{1}{16\pi G}$ with the cosmic evolution. Under such constraint, equation (83) in static background spacetime retaining only linear term in $h_{\mu\nu}$ yields
\begin{equation}
\triangledown^2 h_{00}\simeq \rho_e.
\end{equation}
where, $\rho_e = \rho + \rho_{\mathcal{T}}$ is the effective matter density, $\rho$ and $\rho_{\mathcal{T}}$ being the matter density corresponding to the pressureless dust and tachyonic field respectively. Considering next higher order term in $h_{\mu\nu}$, equation (83) gives

\begin{equation}
\triangledown^2 h_{00} - 3{\mathcal{B}} ~\sqrt{\frac{1}{2}\triangledown^2 h}~\Big(\triangledown^2 h_{00}-\frac{1}{6}\triangledown^2 h\Big) \simeq \rho_e.
\end{equation}
Since at low energy limit Poisson equation is obtained, so Newtonian gravity is valid at weak energy limit. Nevertheless, to pass solar test, an additional constraint $h(\phi) \rightarrow \frac{1}{16\pi G}$ under weak field approximation is required.\\

\section{Concluding remarks}
Earlier attempts to find Noether symmetry for $F(R)$ theory of gravity in vacuum and matter dominated era lead to $F(R) \propto R^{\frac{3}{2}}$ \cite{b1, b2, b3, b4, b5, c}. Such a form is not suitable to explain presently available cosmological data \cite{c}. Search of a better form of $F(R)$ taking a minimally or non-minimally coupled scalar field into account failed to produce symmetry \cite{d}. Apparently, it's not a problem, since not all actions admit symmetry. But then, such a symmetry for $F(R) \propto R^2$ already exists in the literature where an auxiliary variable $Q$ different from $R$ was introduced for the purpose of canonization \cite{e}. Further, $F(R) \propto R^2$ leads to a scale invariant action \cite{sd}. Hence the result should have been reproduced from $F(R)$ theory of gravity. Canonization of the action, treating $R$ as the auxiliary variable creates problem in quantum domain, while canonical formulation of a general $F(R)$ theory of gravity is possible only by treating $R$ as an auxiliary variable, and so we presume that it might be the root of trouble.\\

Earlier, some authors \cite{h} have claimed that $F(R)$ in vacuum admits Noether symmetry for arbitrary power of $R$, if a gauge term is introduced. Such a claim is not true since neither all Noether equations are satisfied nor the conserved current thus obtained satisfies the Hamilton constraint equation \cite{i}. There is yet another more recent claim that Noether symmetry exists for $F(R) \propto R^2$, in the presence of tachyon field \cite{j}. Here, we show that not only the conserved current does not satisfy the field equations, but also the Hamiltonian is not constrained to vanish, which is fundamental of the theory of gravitation. We have also taken up a more general action in section 3 and found that Noether symmetry is not allowed for $F(R)'' \ne 0$. Since all attempts to find Noether symmetry for $F(R)$ theory of gravity (other than the trivial one, viz., $F(R) \propto R^{\frac{3}{2}}$) so far has failed, we conclude that it is not possible to find a form of $F(R)$ by demanding Noether symmetry.\\

To handle $F(R)$ theory of gravity in the presence of Tachyonic field, we have presented a general conserved current in section 5, which is different from one obtained earlier for $F(R)$ being coupled to scalar tensor theory of gravity \cite{d, k, l}. However, one of the conditions required for such conserved current, containing higher order curvature invariant terms, remains unaltered. This part is decoupled from the matter field and may be solved in principle, for the scale factor $a(t)$ independently, choosing a suitable form of $F(R)$ by hand. Thus cosmic evolution remains independent of the choice of the field (Tachyon or scalar field). The conserved current obtained in the process might be an outcome of some higher symmetry, which is not known at present.

\end{document}